\RequirePackage{fix-cm}
\documentclass[smallextended]{svjour3}       
\smartqed 
\usepackage{multirow}
\usepackage{footnote}
\usepackage{tablefootnote}
\usepackage{float}
\usepackage{graphicx}
\usepackage{amsmath}
\usepackage{xcolor}
\usepackage{cite}
\usepackage{multirow}
\usepackage{braket}
\usepackage{amsfonts}
\usepackage{bbold}
\usepackage{subcaption}
\usepackage{support-caption}
\usepackage{subcaption}
\begin{document}

\title{Error--immune quantum communication}
\author{Rajni Bala$^{*}$     \and   V. Ravishankar
}
\institute{ \at Department of Physics, Indian Institute of Technology Delhi, New Delhi-110016, India, \\\email{Rajni.Bala@physics.iitd.ac.in}
}

\date{Received: date / Accepted: date}

\maketitle

\begin{abstract} 
Environmental effects on the transmission of a state result, in general, in a change in the information carried by it. To mitigate this, many techniques such as  quantum error--correcting codes, decoherence--free--subspaces [Rev Mod Phys, 88(4):041001, 2016] are employed. The basic idea underlying them is to protect/recover the state. These techniques require multi-party entanglement, whose generation is a difficult task. Further, retrieval of information  would require complete tomography, which inevitably requires a large number of copies. Taking this into account, in this work, a formalism has been laid down which does not require recovery of a state. The formalism employs scaling laws to obtain quantities that remain invariant under a noisy evolution of a state. The information encoded in these invariant quantities can be transmitted in an error-immune manner. 
 Since multiparty entanglement and error detection/correction will not be required, the proposed scheme would be cost-effective and may be reliably employed for error-free information transfer.
Employing the formalism, we have obtained invariant quantities for various noisy channels of a qubit and a quNit. 
\end{abstract}

\section{Introduction}
\label{intro}
The merger of quantum physics with information theory has resulted in various new applications such as quantum teleportation \cite{teleportation},  superdense coding \cite{Bennett92}, and remote state preparation \cite{RSP}. 
Nonclassical features possessed by quantum states such as superposition, entanglement act as resources in these protocols. 
However, the interaction of quantum states with the environment is inevitable, which results in the corruption of the state, thereby compromising the information carried by it. 

Traditionally, several way-outs to protect a state have been proposed by (i) employing quantum error-correcting codes (QECCs), (ii) searching for decoherence--free--subspaces (DFS), (iii), and introducing  dynamical decoupling (DD) (see \cite{suter2016colloquium} and references therein). In all these approaches, the main task is to preserve/recover a state. Thus, in QECC, information is encoded in a multi-party entangled state which has redundancy inbuilt in it. Detection and correction of possible errors are performed by employing stabilizer measurements and appropriate unitary transformations respectively \cite{devitt2013quantum, nielsen_book, vathsan2015 }. In the DFS technique, information is again encoded in a carefully chosen multipartite entangled state such that subsequent evolution does not corrupt the state \cite{lidar1998decoherence,bourennane2004dfs,cabello2007six}. The third technique, DD coupling, also called bang-bang operations does not require entanglement. However, in this technique, a state is protected from errors by repeated interventions at precisely chosen time intervals. These transformations should be applied fast enough so that the effect of noise on a state can be reversed \cite{byrd2004overview,suter2016colloquium}.

The first two techniques -- QECC and DFS, demand a multipartite entangled state, whose generation is a difficult task \cite{bock2016highly}. The third technique -- DD coupling requires fast controlled operations, that too at fixed time intervals, which are experimentally challenging \cite{byrd2004overview}. More significantly, retrieval of information demands a complete quantum state tomography \cite{paris2004quantum}. This, in turn, implies that a large number of copies of an entangled state are required. In short, information transfer with a single copy of a state is infeasible.

Since multiple copies of a state are inevitably required for information transfer, we may then ask a question: can information be so encoded that it may be transferred and retrieved, with little or no usage of the costly resources such as multipartite entanglement? To answer this question, we take motivation from scaling laws.

Scaling laws provide us with multi--linear quantities that remain invariant for a particular set of phase space trajectories that the system may traverse under certain interactions.
A key example is Kepler's third law and a family of virial theorems.\footnote{The square of the time of revolution $(t)$ in the orbit is as the cube of the semi-major -- axis $(a)$ of the orbit, i.e., $t^2a^{-3}=$ constant.}\cite{LANDAU197613,sun2019classical}. In this work, we adopt this point of view from an information transfer perspective. 
We have laid down a formalism to obtain quantities that remain invariant irrespective of the occurrence of errors. More precisely, these invariants involve combinations of expectation values of the operators that get re-scaled in a compensatory manner under a noisy evolution of a state.  

The invariants may be categorised into three different families. In the first family, invariants are simply the expectation values of operators. The second family consists of ratios of expectation values of two operators. The third family consists of quantities that are combinations of expectation values of operators -- not necessarily scaled by the same factor. 
At this juncture, we wish to point out that invariants which belong to the first family are the ones most studied (see for example \cite{albert2014symmetries}). On the other hand, the invariants belonging to the second and the third family have hitherto not been studied for information transfer, to the best of our knowledge. 

A key advantage of considering the second and the third families (the ratios of expectation values of operators) is the number of independent invariants being significantly larger than that of the first family. Since the ratio of the expectation values of operators cannot be written as the expectation value of the ratio of the operators, these invariants give non--trivial information. In short, formalism equips us with a larger number of independent invariants which can be used to encode the information for error-free transfer.

In the paper, we use formalism to find invariants for various noisy channels of a qubit and a quNit ($N$--level system). The study of quNits is not just of academic interest. For example, quantum communication protocols have been implemented employing orbital angular momentum (OAM) states of light \cite{perumangatt2017quantum,mirhosseini2015high}. Similarly, path and time degrees of freedom using multi-port beam-splitters and pulses in different slots are also employed (see \cite{flamini2018photonic,cozzolino2019high} and references therein). There is an advancement in generation and manipulation of radial quantum number of Laguerre-Gauss states of light and hence may be employed to realise higher-dimensional states \cite{gu2018gouy,fu2018realization,bouchard2018measuring}.  
 Depending on the noisy channels through which these states pass, one may seek to encode information in a way that remains error--free. Note that our information transfer scheme is equally applicable with quantum and classical states of light.

The plan of the paper is as follows: in section (\ref{preliminaries}), a brief discussion of basis operators acting on an $N-$ dimensional state is given. Section (\ref{formalism}) contains the central result of the paper, in which the formalism has been presented to obtain invariants for various noisy channels. The information encoded in these invariants can be transferred without any error. In section (\ref{qubit}), invariants for various noisy channels of a qubit are obtained. Section (\ref{generalization}) discusses generalisation to a quNit, again considering a variety of noisy channels. In section (\ref{robust}), a summary of the results is provided.  
Section (\ref{conclusion}) concludes the paper with closing remarks.
\section{Preliminaries}\label{preliminaries}
In this section, for pedagogic purposes, a  very brief discussion on the basis of an $N-$ dimensional operator space is given.
The set of linear operators acting on an $N$- level system forms a vector space of dimension $N^2$. In a given basis, their representative matrices can be expanded in terms of $N^2$ linearly independent matrices which may be chosen:
\begin{equation}
     \label{eq:off-diagonal}
    S^{(kl)}_{ij}=\delta_{ik}\delta_{jl}+\delta_{il}\delta_{jk},~~A^{(kl)}_{ij}=-i(\delta_{ik}\delta_{jl}-\delta_{il}\delta_{jk}),~~~~d_{ij}^{(k)}=\delta_{ik}\delta_{ij},
  \end{equation}
  where $k,l=\{0,1,\cdots,N-1\}$ and $k>l$ (to aviod double counting).
  The symbols $S^{(kl)}_{ij}$ and $A^{(kl)}_{ij}$ represent $ij^{{\rm th}}$ element of symmetric $(S^{(kl)})$, and anti-symmetric $(A^{(kl)})$  basis matrices respectively. The symbol $ d_{ij}^{(k)}$ represent the $ij^{{\rm th}}$ element of diagonal matrix $ d^{(k)}$. 
 For future use, the diagonal matrices $D^{(kl)}$ are defined as:
\begin{equation}\label{eq:sum_diagonal}
    D^{(kl)}_{ij}=\delta_{ik}\delta_{ij}-\delta_{il}\delta_{ij}= d^{(k)}_{ij}-d^{(l)}_{ij}.
\end{equation}
  For the case of a qubit, the symmetric and anti-symmetric basis operators correspond to Pauli's operators $\sigma_x,\sigma_y$ respectively.

\section{The formalism}
\label{formalism}

In this section, we lay down the formalism to obtain invariants for noisy communication channels, which in turn qualify to be information carriers. 
The underlying formalism of noisy channels is well-established and is given by the so-called completely--positive--trace--preserving (CPTP) maps \cite{nielsen_chuang_2010}.
Here, the action of a noisy communication channel on a state can be represented by a quantum operation $\mathcal{E}(\cdot)$, which is defined by: 
\begin{equation}
    \mathcal{E}(\cdot)=\sum_k E_k(\cdot)E_k^\dagger.
\end{equation}
 The set $\{E_k\}$ is the set of Kraus operators which satisfies the condition $\sum_k E_k^\dagger E_k=\mathbb{1}$.
 
Let $\rho$ be the state of the system used to transfer information. After passing through noisy channel, whose action is well-described by $\mathcal{E}$, the state of the system changes to $\rho'$, 
\begin{equation}
    \rho\rightarrow \rho' = \mathcal{E}\big(\rho\big)= \sum_k E_k\rho E_k^\dagger.
\end{equation}
Given a noisy channel, the idea is to find the operators whose expectation value change in a scaled manner under a noisy evolution of a state. That is, for a given operator $O_\alpha$,
\begin{align}\label{eq:op}
   \langle O_\alpha\rangle_{\rho'} &={\rm Tr}\big(O_\alpha\rho'\big)={\rm Tr}\Big(O_\alpha\sum_k E_k\rho E_k^\dagger\Big)=\sum_k {\rm Tr}\big(E_k^\dagger O_\alpha E_k \rho\big)\nonumber\\
   &=\lambda_\alpha {\rm Tr\big(O_\alpha\rho\big)}=\lambda_\alpha\langle O_\alpha\rangle_{\rho},
\end{align}
where the non-vanishing scaling factor $\lambda_\alpha $ are function of noise parameters. If we demand equation (\ref{eq:op}) to hold for arbitrary states $\rho'$ and $\rho$, the condition assumes the form:
\begin{equation}\label{eq:op1}
    \sum_k E_k^\dagger O_\alpha E_k=\lambda_\alpha O_\alpha.~~~~~~~~~
\end{equation}

The above equation plays an important role in the identification of sets of operators, whose expectation values change in a scaled manner under a noisy evolution of a state. These sets of operators are employed to obtain quantities that remain invariant and hence, can be used for encoding information.
\subsection{Identification of invariants}
The equation (\ref{eq:op1}) provides us with a set of operators whose expectation values change in a scaled manner. The scaling with which these expectation values change is a function of noise parameters. Therefore, the task of finding invariants reduces to finding a function of expectation values of these operators such that it is independent of any noise parameters.

To obtain this, consider the set of operators  $\{O_1,O_2,\cdots,O_n\}$ that satisfy equation (\ref{eq:op1}), i.e.,

\begin{align}
    &\sum_k E_k^\dagger O_1 E_k=\lambda_1 O_1,\quad\quad\quad \sum_k E_k^\dagger O_2 E_k=\lambda_2 O_2,\nonumber\\
    &\sum_k E_k^\dagger O_3 E_k=\lambda_3 O_3,~~\cdots , ~~\sum_k E_k^\dagger O_n E_k=\lambda_n O_n,
\end{align}
where $\lambda_1,~\lambda_2,\cdots ,~\lambda_n$ represent the scaling with which the expectation values of respective operators $O_1,~O_2,\cdots,~O_n$ change under the noisy evolution of a state. That is to say,
\begin{align}\label{eq:o1}
   & \langle O_1\rangle_{\rho'}=\lambda_1\langle O_1\rangle_{\rho}~,\quad\quad\quad ~~~~ \langle O_2\rangle_{\rho'}=\lambda_2\langle O_2\rangle_{\rho}~,\nonumber\\
    &\langle O_3\rangle_{\rho'}=\lambda_3\langle O_3\rangle_{\rho}~,~~\cdots, ~~~~~\langle O_n\rangle_{\rho'}=\lambda_n\langle O_n\rangle_{\rho}~.
\end{align}
Consider the following function of expectation values,
\begin{equation}\label{eq:product}
 \prod_{\alpha=1}^n\bigg( \langle O_\alpha\rangle_{\rho'}\bigg)^{r_\alpha}=\prod_{\alpha=1}^n \lambda_\alpha^{r_\alpha}\bigg( \langle O_\alpha\rangle_{\rho}\bigg)^{r_\alpha},
\end{equation}
where $r_\alpha$ can take real values.
 For a given set $\{\lambda_1,\lambda_2,\cdots,\lambda_n\}$, if there exists a set of values $\{r_1,r_2,\cdots,r_n\}$, for which 
\begin{equation}\label{eq:invariant_condition}
    \prod_{\alpha=1}^n \lambda_\alpha^{r_\alpha}=1,
\end{equation} 
then, the quantity $\prod_{\alpha=1}^n \langle O_\alpha\rangle^{r_\alpha}$ is invariant.
 
 We now categorise these invariants into three different families depending upon how the operators are re-scaled under a noisy evolution of a state. 

 
\subsubsection*{The first family of invariants }
The first family of invariants corresponds to the situation when there is only one operator, i.e., $n=1$ and hence, $\lambda = 1$. Then, the condition (\ref{eq:op1}) reduces to: 
\begin{equation}\label{eq:operator}
    \sum_k E_k^\dagger O E_k= O .
\end{equation}
 This condition implies that the expectation value of operator $O$ remains unchanged under a noisy evolution of a state, and hence, can be used for encoding information. If there are $k$ such linearly independent operators $\{O_k\}$, then, corresponding invariants are denoted by $\mathcal{I}^{(k)}=\langle O_k\rangle$. 
This particular set of invariant quantities is also obtained for Lindblad master equations and is used to recognize steady-state structures in a noisy state \cite{albert2014symmetries}.

 
 \subsubsection*{The second family of invariants} \label{second family}
 There may be instances in which no invariant can be obtained from the first family. In fact, the invariants provided by the first family are rather limited in number. Therefore, we consider a more general case and define the second family of invariants. In the second family, the ratio of expectation values of the two operators acts as an invariant. That is, for a pair of operators $O_1,$ and$~O_2$, for which $\lambda_1=\lambda_2 =\lambda\neq 1$, the expectation values of the two operators follow the relation, 
    \begin{equation}
    \label{eq:trace}
   \langle O_1\rangle_{\rho'}=\lambda \langle O_1\rangle_{\rho}~,~~~~~~~~~\langle O_2\rangle_{\rho'}=\lambda \langle O_2\rangle_{\rho}~,
\end{equation}
under the noisy evolution of a state.
  For these operators, equation (\ref{eq:invariant_condition}) is satisfied for $r_1=-r_2=1$. This implies that the information encoded in the quantity,

\begin{equation}\label{eq:operator2}
    \mathcal{I}=\frac{\langle O_1\rangle}{\langle O_2\rangle},
\end{equation} 
 remains free from errors and can be transferred reliably. 
The set of linearly independent operators satisfying the above condition constitutes the second family of invariants. 
\subsubsection*{The third family of invariants}\label{third family}
We have, so far, discussed the two families of invariants. 
Though together these two families provide a number of invariants, they do not exhaust all possible independent invariants. There may exist a particular set of operators which satisfies the condition (\ref{eq:invariant_condition}), but does not belong to either of the above-mentioned families. Therefore, we define the third family of invariants. The third family of invariants consists of those functions for which the condition (\ref{eq:invariant_condition}) is satisfied either for $n>2$ or for $n=2$ when $r_1\neq -r_2=1$ .


Thus, the three families of invariants, taken together provide us with a number of independent invariants for a given noisy channel. 


In the subsequent sections, employing the formalism, we obtain invariants for various noisy channels of a qubit and a quNit.


\section{Information transfer with qubits}
\label{qubit}
In this section, prominent channels such as bit-flip, phase-flip, amplitude-damping are discussed. Owing to their effect on real-world implementation of communication protocols, these have been studied for quantum teleportation in \cite{fortes2015fighting}.    
Many of these noisy channels are special cases of the Pauli channel \cite{nielsen_chuang_2010} in which Kraus operators are unitary operators with appropriate weights. Therefore, we first discuss this case of Kraus operators.

For a given noisy channel, let the set of Kraus operators be scaled unitary operators, $E_k=\sqrt{p_k}U_k$, where the associated weights $p_k$ satisfy $\sum_k p_k=1$, and $U_k$ is an unitary operator. For such a channel, if there exists an operator $O$ that commutes with all the unitary operators, i.e., $[O,U_k]=0 $, then, the condition $\sum_k E_k^\dagger O E_k =O$ is automatically satisfied. This implies that the expectation value of operator $O$ remains unchanged and hence belongs to the first family of invariants. 

In a Pauli channel, bit-flip, phase-flip and a combination of both can corrupt the system, that too with arbitrary probabilities. Thus, a Pauli channel is characterised by three independent parameters. 
A state $\rho$, after passing through a Pauli channel, changes to:
\begin{equation}\label{eq:Pauli}
    \rho\rightarrow \rho'= p_0\rho
+p_1 \sigma_x\rho  \sigma_x+p_2  \sigma_y\rho  \sigma_y +p_3  \sigma_z \sigma_z~,~~~~  
\end{equation}
where $\{p_0,p_1,p_2,p_3\}$ are statistical weights associated with each operators $\{\mathbb{1},~\sigma_x,~\sigma_y,~\sigma_z\}$.
Such a general channel does not admit any invariant except trivial identity. Hence, we consider various noisy channels, a few of which are special cases of the Pauli channel, and obtain invariants for error-free information transfer.



\subsection{Bit-flip channel} 
\label{bit_qubit}
A probabilistic bit-flip channel can be modeled with Kraus operators $\mathbb{1}$ and $\sigma_x$, each occurring with a respective probability of $(1-p)$ and $p$. This is a special case of Pauli channel when $p_1=p$ and $p_2,~p_3=0$.
A state $\rho$ after passing through such channel changes to $\rho'$ \cite{nielsen_chuang_2010}:
\begin{equation}\label{eq:bit}
\rho'= (1-p)\rho +p~ \sigma_x\rho  \sigma_x.
\end{equation}
This channel admits two invariants. They are,
\begin{align}
 &\mathcal{I}_1=\langle  \sigma_x\rangle,~~~\mathcal{I}_2=\frac{\langle  \sigma_y\rangle}{\langle  \sigma_z\rangle}.
\end{align}
 Note that $\mathcal{I}_1,$ and$~\mathcal{I}_2$
 belong to the first and the second family of invariants respectively. Details of demonstration of invariance of these quantities are given in the appendix (\ref{app:bit-flip}).

Such channel is studied for polarization states of a photon in \cite{dong2014}. The state to be transferred is encoded in a non-maximally entangled bi-partite polarized state. 

\subsection{Phase-flip channel}
\label{phase_qubit}
Phase flip channel is a special case of Pauli channel in which Kraus operators are $\mathbb{1}$ and $\sigma_z$ which act on a state with a respective probability of $(1-p)$ and $p$. A State $\rho$, after passing through the channel changes to state $\rho'$ \cite{nielsen_chuang_2010}:
\begin{equation}
    \rho\rightarrow \rho'= (1-p)\rho + p~ \sigma_z\rho~  \sigma_z.
\end{equation}
This channel also admits two invariants,
\begin{align}
&\mathcal{I}_1=\langle  \sigma_z\rangle,~~~~~\mathcal{I}_2=~\frac{\langle  \sigma_x\rangle}{\langle  \sigma_y\rangle},
\end{align}
belonging to the first and the second family respectively.
 The information encoded in the invariant $\langle\sigma_z\rangle$ remains unchanged because the phase-flip error does not change the relative population in its eigenstates and thus acts as an invariant. The phase-flip operator $\sigma_z$ flips the eigenstates of the operators $\sigma_x$ and $\sigma_y$ (i.e., the eigenstate  with eigenvalue $+1$ is changed to the state with eigenvalue $-1$ and vice-versa). Due to this, the expectation values of both the operators change in a commensurate manner, and thus the ratio of the two acts as another invariant. Since $\mathcal{I}_2$ involves the ratio of expectation values of two operators which may be small, suitable sensitivity of the detector is required.

\subsection{ Combined bit and phase-flip channel} \label{both_qubit}
This channel is modeled by unitary operators $\mathbb{1}$ and $\sigma_y$. 
A state $\rho$ after passing through this channel changes to \cite{nielsen_chuang_2010}:
\begin{equation}
    \rho\rightarrow \rho'= (1-p)~\rho + p \sigma_y\rho~ \sigma_y.
\end{equation}
This channel admits two invariants, 
\begin{equation}
\mathcal{I}_1=\langle  \sigma_y\rangle,~~~~~~\mathcal{I}_2=~\frac{\langle  \sigma_x\rangle}{\langle  \sigma_z\rangle},
\end{equation}
belonging to the first and the second family of invariants. Since Kraus operators involved are $\mathbb{1}$ and $\sigma_y$, they have no effect on the eigenstates of $\sigma_y$ whereas eigenstates of $\sigma_x$ and $\sigma_z$ are flipped. This results in the commensurate change in the expectation values of the two and hence, ratio of the two is invariant.  

\subsection{Dephasing channel}\label{dephaisng_qubit}
A dephasing channel is used to model a random phase acquired by scattering of photons. Such a channel occurs in polarization and OAM states of light passing through fiber which have been studied in \cite{gupta2015preserving,gupta2016dephasing} using standard techniques.  A state $\rho$ after passing through this channel changes to \cite{nielsen_chuang_2010},
\begin{equation}
    \rho\rightarrow\rho' =\mathcal{E}(\rho)= \alpha \rho + (1-\alpha)  \sigma_z\rho~  \sigma_z,
\end{equation}
 where $\alpha= \frac{1+\sqrt{1-\lambda}}{2}$, $\lambda$ is the phase damping parameter. The effect of this channel on a state is to reduce coherence of the state. The two invariants,
 \begin{equation}\label{eq:inv_dephase}
     \mathcal{I}_1= \langle  \sigma_z\rangle ,~~~~\mathcal{I}_2= \frac{\langle  \sigma_x\rangle}{\langle  \sigma_y\rangle},
 \end{equation}
 belong to the first and the second family respectively.
 The information encoded in these quantities can be transferred in an error-immune manner. The dephasing channel for a qubit has also been studied in \cite{albert2014symmetries} in which conserved quantity $ \mathcal{I}_1$ is obtained.


\subsection{Depolarizing channel}
\label{depolar_qubit}
A depolarizing channel studies the effect of multiple random scattering of the signal \cite{soorat2014noise}. This multiple random scattering leads to the loss of information. This is equivalent to the incoherent addition of a maximally mixed state (white noise) to the signal with some probability $p$. 
Therefore, a state $\rho$ carrying the information after passing through this channel changes to \cite{nielsen_chuang_2010}:
\begin{equation}
\rho'= (1-p)\rho+\dfrac{p}{2}\mathbb{1}.
\end{equation}
For this noisy channel, there exists no operator whose expectation value remains invariant, i.e., no invariant from the first family exists. However, the expectation values of all the three operators $ \sigma_x,~ \sigma_y,$ and $ \sigma_z$ acquire the same multiplicative factor in accordance with equation (\ref{eq:trace}). Therefore, the ratio of the expectation values of any two operators remains unchanged and hence can be used to encode information.
  That is to say, information encoded in the two invariants,
\begin{align}
 &\mathcal{I}_{1}=\frac{\langle  \sigma_x\rangle}{\langle  \sigma_z\rangle},~~~~~\mathcal{I}_{2}=\frac{\langle  \sigma_y\rangle}{\langle  \sigma_z\rangle},
\end{align}
can be transferred without any errors even for a very large amount of noise. 
\subsection{Equiprobable bit-flip and phase-flip noisy channel}\label{equi_qubit}
In the above channel, the probabilities of occurrences of all three errors are the same. However, a more general situation corresponds to the case when only two of the three probabilities are the same. To be specific, consider that probabilities of bit-flip and phase-flip are the same, i.e., both errors corrupt a state with the same probability $p_1$. A state $\rho$ after passing through this channel can be written as:
\begin{equation}
   \rho'= p_0\rho 
+p_1 ( \sigma_x\rho~  \sigma_x+\sigma_z\rho~  \sigma_z)+p_2  \sigma_y\rho~  \sigma_y.
\end{equation} 
In this case, the expectation values of both the observables $ \sigma_x$ and $ \sigma_z$ get reduced by the same multiplicative factor which is a function of probabilities of error. Thus, information encoded in the ratio of the two,
\begin{equation}
    \mathcal{I}=\frac{\langle  \sigma_x\rangle}{\langle  \sigma_z\rangle},
\end{equation}
remain free from errors and hence can be transferred reliably.

In this case, we have obtained an invariant explicitly for the case when probabilities for bit-flip and phase-flip are the same. However, it admits a straightforward generalisation for noisy channels when any of the two probabilities of errors are the same, i.e., $p_i=p_j$ for some $\sigma_i,~ \sigma_j$ where\footnote{$\sigma_1\equiv \sigma_x,~\sigma_2\equiv\sigma_y,$ and $\sigma_3\equiv\sigma_z$.} $i,j=\{1,2,3\}$. The quantity  $\mathcal{I}=\frac{\langle \sigma_i\rangle}{\langle\sigma_j\rangle}$ is invariant, and hence can be used for encoding information.

\subsection{Amplitude damping channel (ADC)}\label{AD}
This channel does not fall under the category of the Pauli channel. It was originally introduced to study spontaneous emission. In an optical communication system, this channel is employed to study the effect of scattering and attenuation on the system. In the amplitude--damping channel, the population in one level depletes, which results in a decrease in the coherence of a state. The loss of depletion of the population in one level is compensated by an increase of population at another level. The relevant Kraus operators which mimick the action of such noise are given as\cite{nielsen_chuang_2010}:
\begin{equation}
E_0=\begin{pmatrix}
1&0\\
0& \sqrt{1-q}
\end{pmatrix},\quad
E_1=\begin{pmatrix}
0&\sqrt{q}\\
0& 0
\end{pmatrix};\quad q\in [0,1].
\end{equation}
A state $\rho$, after passing through this channel changes to state $\rho'$ :
%
 \begin{equation}\label{eq:ad}
 \rho=\begin{pmatrix}
 \rho_{00}& \rho_{01}\\
 \rho_{10}& \rho_{11}\end{pmatrix}\rightarrow \rho'=\begin{pmatrix}
 \rho_{00}+q\rho_{11} & ~~~\sqrt{1-q}\rho_{01}\\
 \sqrt{1-q}\rho_{10}&~~~~~(1-q)\rho_{11}
 \end{pmatrix}. \end{equation}
The two invariants, belonging to the respective second and third family are,
\begin{equation}
    \mathcal{I}_{1}~=~\dfrac{\langle  \sigma_x\rangle}{\langle  \sigma_y\rangle},~~~~
\mathcal{I}_{2}~=~\dfrac{\langle  \sigma_x\rangle\langle  \sigma_y\rangle}{\langle\pi_z^-\rangle},
\end{equation}
where $\pi_z^-$ is the projection operator of $\sigma_z$ with eigenvalue $-1$.

Thus, if the channel through which the state gets transmitted can be modeled as an amplitude damping channel, error-free information can be transferred by encoding in the two invariants. 


\subsection{Generalised amplitude--damping channel (GADC)} \label{amp_qubit}
The generalised amplitude damping channel describes the process in which the population of both levels leaks to each other. As a result, coherence in a state decreases. In an optical communication system, this channel is employed to study the effect of scattering and attenuation on the system \cite{nielsen_chuang_2010}.

Let $\rho$ be the state used to transfer the information. After passing through the channel, the state changes to:
\begin{equation}
    \rho\rightarrow\rho'= \sum_{j=1}^4 E_j\rho E_j^\dagger ,
\end{equation}
where the representative Kraus operators are given by \cite{nielsen_chuang_2010} :
\begin{align}
    &E_1=\sqrt{p_1}\begin{pmatrix}
1&0\\
0& \sqrt{1-q}
\end{pmatrix},\quad
E_2=\sqrt{p_1}\begin{pmatrix}
0&\sqrt{q}\\
0& 0
\end{pmatrix};\quad\quad q\in [0,1],\nonumber\\
&E_3=\sqrt{p_2}\begin{pmatrix}
\sqrt{1-q}&0\\
0& 1
\end{pmatrix},\quad
E_4=\sqrt{p_2}\begin{pmatrix}
0&0\\
\sqrt{q}& 0
\end{pmatrix};\quad \quad p_1+p_2=1.
\end{align}

The noisy state $\rho'$ is given as:
 \begin{equation}\label{eq:gad}
  \rho'=\begin{pmatrix}
\big( p_1+p_2(1-q)\big)\rho_{00}+p_1q \rho_{11}& \sqrt{1-q}\rho_{01}\\
 \sqrt{1-q}\rho_{10}&~~~~~\big(p_1(1-q)+p_2\big)\rho_{11}+p_2q\rho_{00}
 \end{pmatrix} \end{equation}
 
 This channel admits only one invariant,
 \begin{equation}
    \mathcal{I}=\dfrac{\langle  \sigma_x\rangle}{\langle  \sigma_y\rangle},
\end{equation}
which belongs to the second family. The information encoded in the invariant can be transferred without any errors. 


 \section{Information transfer with quNits}
\label{generalization}
 It is a well-known fact that higher-dimensional states carry a larger amount of information. Owing to this, these states are used extensively in communication protocols. 
Therefore, in this section, invariants are obtained for various noisy channels of a quNit for error-free information transfer.

Similar to the case of a qubit, many channels of a quNit are special cases of what is known as the generalised Pauli channel (GPC) \cite{miller2018propagation}. A generalised Pauli channel is the one in which an error is generated by the generalised Pauli operators each occurring with some finite probability. The generalised Pauli operators on a single quNit are products of powers of operators $X$ and $Z$ which are defined as:
\begin{equation}\label{eq:operator_X}
    X=\sum_{k=0}^{N-1} \ket{k+1}\bra{k},\quad\quad Z= \sum_{k=0}^{N-1} \omega^k \ket{k}\bra{k} ,
\end{equation}
where $\omega= e^{\frac{2\pi i}{N}}$ and addition of integers is considered modulo $N$.  Both the operators $X$ and $Z$ are unitary operators satisfying $X^N,~Z^N=\mathbb{1}$.

The action of the GPC on a state $\rho$ may be expressed as:
\begin{equation}\label{eq:GP}
    \rho\rightarrow\rho'=\sum_{r,s=0}^{N-1} ~p_{rs} (X^rZ^s)~\rho ~(X^rZ^s)^\dagger,~~~~~~~\sum_{r,s=0}^{N-1} p_{rs}=1,
\end{equation}
where $0\leq p_{r,s}\leq 1$ is the probability with which the unitary operator $X^rZ^s$ corrupts the state.
The GPC admits no invariants.

We now consider various noisy channels of a quNit, a few of them are special cases of the GPC. The sets of invariants obtained can be employed for error-free information transfer.
\subsection{Generalised flip error}
\label{quNitflip}

In a generalised flip error, all the relevant Kraus operators are generated by operator $X$. 
 Let $\rho$ be the state used to send information through such channel. After passing through this channel, it changes to the state $\rho'$:
 \begin{equation}\label{eq:quditflip}
     \rho\rightarrow\rho'= \sum_{r=0}^{N-1} p_r X^r\rho~ (X^r)^\dagger .
 \end{equation}
  One may immediately say that the expectation values of operators, $\mathcal{I}_{1}^{(m)}=\langle X^m\rangle$ is invariant, and thus encoded information remain error--free. In addition, the second family of invariants provides us with the two sets,
 \begin{equation}
\mathcal{I}_{2}^{(m)}=\frac{\langle Z^m\rangle}{\langle (XZ)^m\rangle},~~~\mathcal{I}_{3}^{(ml)}=\frac{\langle Z^m\rangle}{\langle Z^mX^l\rangle},~~~~m\neq l,
 \end{equation}
 which remain immune to errors and can be transferred reliably. The demonstration of invariance of these quantities is given in the appendix (\ref{app:quNit_flip}).
 The operators appearing in the three sets of invariants are unitary and hence, each quantity gives the two real invariants.

 Whereas the sets $\mathcal{I}_{1}^{(m)}$ and $\mathcal{I}_{2}^{(m)}$ each provide $(N-1)$ independent quantities, the third set $\mathcal{I}_{3}^{(ml)}$ provides $(N-1)(N-2)$ independent quantities which is quadratic in $N$. In all, information encoded in $(N^2-N)$ independent invariants can be transferred in an error-immune manner. 

\subsection{Generalised phase error}\label{quNitphase}
Generalised phase error channel is also a special case of generalised Pauli channel if we put $r=0$. This clearly implies that all the relevant Kraus operators of this channel are generated by operator $Z$.
Let $\rho$ be a state used for transferring information. After passing through such channel, it changes to:
\begin{equation}
    \rho\rightarrow\rho'=\sum_{s=0}^{N-1} p_s Z^s\rho ~(Z^s)^\dagger, 
\end{equation} 
where each Kraus operator $Z^s$ corrupts the state with probability $p_s$. The set of quantities, $\mathcal{I}_{1}^{(m)}=\langle Z^m\rangle$, is invariant and belongs to the first family. Additionally, the two sets of quantities, belonging to the second family of invariants, are
\begin{equation}
     \mathcal{I}_{2}^{(m)}=\frac{\langle X^m\rangle}{\langle (XZ)^m\rangle},~~~~~    \mathcal{I}_{3}^{(ml)}=\frac{\langle X^m\rangle}{\langle X^mZ^l\rangle},~~~~~~m\neq l.
\end{equation}
Whereas the two sets $\mathcal{I}_{1}^{(m)}$ and $\mathcal{I}_{2}^{(m)}$ contain $(N-1)$ independent quantities each, the third set $\mathcal{I}_{3}^{(ml)}$ contain $(N-1)(N-2)$ independent quantities. Thus, information encoded in $(N^2-N)$ independent quantities can be transferred in an error-immune manner. 
\subsection{Generalised combined flip and phase error}\label{quNitboth}
Higher-dimensional generalisations of flip and phase errors of a qubit are represented by $X$ and $Z$ respectively. In a similar way, their product $XZ$ corresponds to the higher-dimensional generalisation of operator $\sigma_x\sigma_z$ of a qubit.
Therefore, in a generalised combined flip and phase error, all the Kraus operator are generated by operator $XZ$. A state $\rho$ after passing through such channel changes to:
\begin{equation}
    \rho\rightarrow\rho'= \sum_{r=0}^{N-1} p_r(XZ)^r\rho~ \Big((XZ)^r\Big)^\dagger ,
\end{equation}
which is a special case of a generalised Pauli channel if we put $r=s$. The four sets of invariants are,
\begin{eqnarray}
    &&\mathcal{I}_{1}^{(m)}= \langle (XZ)^m\rangle,~~~~~\mathcal{I}_2^{(m)}=\frac{\langle X^m\rangle}{\langle Z^{N-m}\rangle},\nonumber\\
    &&\mathcal{I}_{3}^{(ml)}=\frac{\langle X^m\rangle}{\langle X^lZ^{l-m}\rangle},~~~\mathcal{I}_{4}^{(ml)}=\frac{\langle Z^m\rangle}{\langle X^{l-m}Z^l\rangle},~~~~~~~~~m\neq l.
\end{eqnarray}
The sets $\mathcal{I}_{1}^{(m)}$ and $\mathcal{I}_{2}^{(m)}$, belonging to the first and the second family of invariants respectively, contain $(N-1)$ independent quantities each.  Whereas the sets $\mathcal{I}_{3}^{(ml)}$ and $\mathcal{I}_{4}^{(ml)}$ contain $\binom{N-1}{2}$ independent quantities each which is quadratic in $N$. In all, information encoded in $(N^2-N)$ independent quantities can be transferred in an error-immune manner. 
 
 \subsection{Flip error considering all possible transpositions}
\begin{center}
    \begin{figure}[!htb]
\centerline{\includegraphics[scale=0.3]{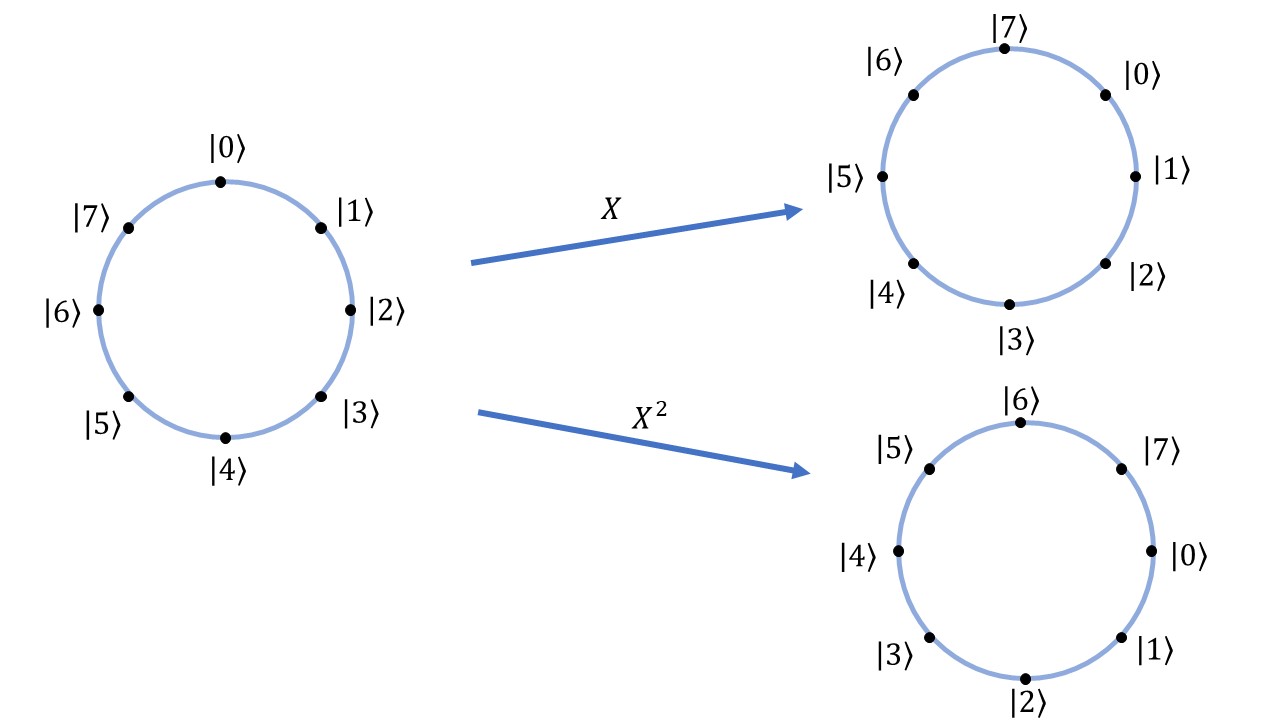}}
\caption{Pictorial representation of the action of flip operators $X$ and $X^2$. The action of operator $X$ is to shift the position of each state to the next one in a clockwise direction. The operator $X^2$ shifts each state to the second next position.}
\label{fig:circle}
\end{figure}
\end{center}

\begin{center}
    \begin{figure}[!htb]
\centerline{\includegraphics[scale=0.5]{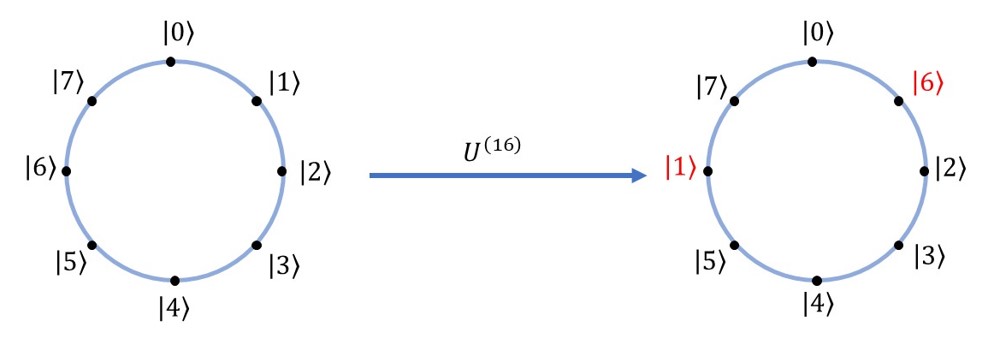}}
\caption{Pictorial representation of action of transposition $U^{(16)}$ on the basis states $\{\ket{i}\}$. Here, we have considered an eight-dimensional state space for illustration purposes. The equidistant points on the circle represent the states. The states marked in red are flipped into each other.}
\label{fig:transposition}
\end{figure}
\end{center}

  In  subsection (\ref{quNitflip}), we have studied generalised flip errors in which different errors were introduced by repeated application of operator $X$ defined in equation (\ref{eq:operator_X}). If all the states of an $N$-- level system are considered as $N$ equidistant points on a circle, then, the effect of error introduced by operator $X$ is to shift each state to the next position in a cyclic order. Similarly, the effect of $X^2$ is to shift each state to the second next position in a cyclic order.  Thus, the assumption that each error occur with different probabilities was completely legitimate. The effect of these operators is shown in the Fig. \ref{fig:circle} for an eight--dimensional system.
  However, these are not the only flip errors that can corrupt an $N$--level system.  There may be cases in which only the two states are flipped into each other leaving all the other states unchanged. These errors are mere transpositions. The effect of transposition on an eight--dimensional state is shown in Fig. \ref{fig:transposition}. For an $N$-- level system, there can be in total $\binom{N}{2}$ possible transpositions. The probability $(p)$ of occurrence of any of these errors is the same as there is no prior reason why one transposition should occur with a higher probability than another.  
  Let $U^{(mn)}$ be the unitary operator which represents the transposition of the two states $\ket{m} $ and $\ket{n}$. It is given by 
  \begin{equation}
      U^{(mn)}= \ket{m}\bra{n}+\ket{n}\bra{m}+\sum_{\substack{j=0,\\ j\neq m,n}}^{N-1} \ket{j}\bra{j},\quad \quad m< n.
  \end{equation}
   A state $\rho$, after passing through this channel, changes to,
  \begin{equation}
    \rho\rightarrow\rho'= (1-p)\rho +p\sum_{\substack{m,n=0,\\ m< n}}^{N-1} U^{(mn)}\rho~\big(U^{(mn)}\big)^\dagger. 
  \end{equation}
  There is only one operator which commutes with all the operators $U^{(mn)}$ and may identified to be:
  \begin{equation}
      O=\sum_{\substack{k,l=0,\\ k>l}}^{N-1}  S^{(kl)}.
  \end{equation}
  Therefore, in accordance with equation (\ref{eq:operator}), the expectation value of operator, $\mathcal{I}_1=\langle O\rangle$, remains unchanged, and hence encoded information can be transferred without any error. 
  
  To identify additional invariants, we first define two sets of operators,
  \begin{equation}
     O_2^{(k)}= \sum_{l=0}^{N-1} A^{(kl)},~~~~O_3^{(k)}=\sum_{\substack{l=0}}^{N-1} \Big(S^{(kl)}  -  S^{(N-1-k~l)}\Big);~~~~~k\neq l,~N-1-k\neq l.
  \end{equation}
 One may check that there are three sets of invariants namely, 
  \begin{align}
      &\mathcal{I}^{(k)}_{2}=\dfrac{\langle O_2^{(k)}\rangle}{\langle D^{(k~N-1)}\rangle},~~~~\mathcal{I}^{(k)}_{3}=\dfrac{\langle O_3^{(k)}\rangle)}{\langle D^{(k~N-1)}\rangle},\quad\quad\mathcal{I}^{(kl)}_{4}=\dfrac{\langle D^{(kl)}\rangle}{\langle D^{(k~N-1)}\rangle},
  \end{align}
  all belonging to the second family.
  The number of invariants for the sets $\mathcal{I}^{(k)}_{2},~\mathcal{I}^{(k)}_{3}$ and $\mathcal{I}^{(kl)}_{4}$ are $(N-1)$ and $[\frac{N}{2}]$ and $(N-2)$, where $[x]$ gives the integer part of $x$. In total, information encoded in $([\frac{5N}{2}]-2)$ independent quantities can be transferred in an error-immune manner.
 \subsection{Dephasing channel}
In a dephasing channel, the coherence of a state decreases without any change in the population of a state.
Let $\rho$ be the state of a quNit system used to transfer information. After passing through dephasing channel, it changes to \cite{marques2015experimental}:
\begin{equation}
    \rho\rightarrow\rho'=\sum_{j=0}^N p_j ~E_j\rho E_j^\dagger ,
\end{equation}
where the relevant Kraus operators are $E_j =\mathbb{1}-2\ket{j}\bra{j}$ and $E_N=\mathbb{1}$.
The invariants in which information can be encoded for error-free transmission are:
\begin{equation}
    \label{eq:inv_quNit_dephase}
    \mathcal{I}^{(k)}_{1}=\langle D^{(kk+1)}\rangle,~~~~~~\mathcal{I}_{2}^{(kl)}=\frac{\langle S^{(kl)}\rangle}{\langle A^{(kl)}\rangle},~~~~k>l.
\end{equation}
Thus, in the dephasing channel, information encoded in $\big(\frac{N(N+1)}{2}-1\big)$ invariants can be transferred in an error-immune manner.
 \subsection{Depolarising channel}
\label{depolar_qudit}

The effect of depolarizing channel on the state is to incoherently mix the state with white noise with probability $p$, as has also been seen in section (\ref{depolar_qubit}). The effect of this channel on a quNit is similar to that on a qubit.

Let $\rho$ be the state of a quNit used to transmit information through depolarizing channel. After passing through depolarizing channel, the state of the system changes to \cite{wilde_2017}: \begin{equation}
\rho\rightarrow\rho'=(1-p)\rho +\dfrac{p}{N}\mathbb{1}.
\end{equation} 
The invariants, in which information can be encoded for error-free transmission are, \begin{eqnarray}
        &&\mathcal{I}_{1}^{(kl)}=\dfrac{\langle S^{(kl)}\rangle}{\langle D^{(kl)}\rangle},~~~~~~\mathcal{I}_{2}^{(kl)}=\dfrac{\langle A^{(kl)}\rangle}{\langle D^{(kl)}\rangle},~~~~~~~~~~~~~~~k>l,\\
 && \mathcal{I}_{3}^{(mnl)}=\dfrac{\langle D^{(mn)}\rangle}{\langle D^{(ml)}\rangle},~~~~~~~~~~~~~~~~~~~~~~~~~~~m\neq n \neq l.
\end{eqnarray}  

The first two sets contain $\binom{N}{2}$ independent invariants, while the third set contains $(N-2)$ independent invariants. Thus, in the depolarizing channel, for an $N-$ dimensional system, information encoded in $(N^2-2)$ invariants can be transferred reliably.

\subsection{Amplitude damping channel (ADC)}
In this section, we obtain a set of invariants for a quNit as a straightforward generalisation of a qubit case.

Let $\rho$ be the state of a quNit used to transfer information. After passing through the amplitude damping channel, the state changes to:
\begin{equation}
    \rho\rightarrow\rho'=\mathcal{E}(\rho)= E_0\rho E_0^\dagger +\sum_{m,n=1}^{N-1} E_{mn}\rho E_{mn}^\dagger.
\end{equation}
The relevant Kraus operators are given as \cite{chessa2021quantum}: 
\begin{align}
    &E_0= \ket{0}\bra{0}+\sum_{n=1}^{N-1} \sqrt{1-\xi_n}\ket{n}\bra{n},~~~~~~~~~\xi_n=\sum_{0\leq m <n}\gamma_{nm},\nonumber\\
    &E_{mn}=\sqrt{\gamma_{nm}}\ket{m}\bra{n};~~~~~\forall ~ m,n~ {\rm st} ~0\leq m<n\leq N-1,
\end{align}
where $\gamma_{nm}$ describe the rate with which population from the $n^{{\rm th}}$ level is transferred to the $m^{{\rm th}}$ level.
The set of invariants for this channel are given by the second  and the third family as follows:
\begin{equation}
    \mathcal{I}^{(kl)}_{1}= \frac{\langle S^{(kl)}\rangle}{\langle A^{(kl)}\rangle}, ~ ~~k>l;~~~~ ~~~~~~
    \mathcal{I}_{2}= 
    \frac{\langle S^{(N-1~0)}\rangle\langle A^{(N-1~0)}\rangle}{\langle\pi_{N-1}\rangle},
\end{equation}
where the symbol $\pi_{N-1}$ represents the projection operator for the state $\ket{N-1}$.
The set $\mathcal{I}_1^{(kl)}$ contains $\binom{N}{2}$ independent quantities.
This implies information encoded in these $\big(\binom{N}{2}+1\big)$ invariants can be transferred without any errors.

\subsection{Generalised amplitude damping channel (GADC)}\label{quNitGAD}
In section (\ref{amp_qubit}), we have discussed invariants for generalised amplitude damping channels for a two-level system.  For an $N-$ dimensional system, the generalisation is straightforward. 

Let $\rho$ be the state of a quNit used to transfer the information encoded in it. After passing through a generalised amplitude damping channel, the state of the system can be written as:
\begin{equation}
    \rho\rightarrow\rho'=\mathcal{E}(\rho)= \sum_{m=0}^{N-1} E_m\rho E_m^\dagger+\sum_{m,n=0}^{N-1} E_{mn}\rho E_{mn}^\dagger,
\end{equation}
where the Kraus operators $E_m,~E_{mn}$ mimick the action of channel and are given as:
\begin{align}
   & E_{m}=\sqrt{p_m}\bigg(\sqrt{1-\xi_m} \ket{m}\bra{m}+\sum_{n=0,\newline n\neq m}^{N-1}\ket{n}\bra{n}\bigg)\\
    &E_{mn}=\sqrt{p_m}\sqrt{\gamma_{nm}}\ket{n}\bra{m},\quad\quad \xi_m=\sum_{n=0,\newline n\neq m}^{N-1}\gamma_{nm}
\end{align}
where $\gamma_{nm}$ represent the rate at which population of $m^{\rm th}$ level falls to $n^{\rm th}$ level. The symbol $p_m$ represents the weight of the Kraus operator.
The invariant for this channel, in accordance with equation (\ref{eq:operator2}), is the ratio of expectation values of $S^{(kl)}$ and $A^{(kl)}$, i.e.,
\begin{equation}
    \mathcal{I}^{(kl)}=\frac{\langle S^{(kl)}\rangle}{\langle A^{(kl)}\rangle},~~~~~~~~~~~~~~k>l.
\end{equation}
Thus, in the generalised amplitude damping channel, information can be transferred in an error-immune fashion by encoding in these $\binom{N}{2}$ independent invariants.

\section{Summary of results}\label{robust}

In this section, we summarise the results obtained for various noisy channels of a qubit and a quNit. 
The formalism proposed in this paper provides three families of invariants -- (i) in which information is encoded in the expectation value of the operators, (ii) in which information is encoded in the ratio of expectation values of the two operators, (iii) in which information is encoded in combinations of expectation values of operators -- not scaled by the same factor. 




\begin{table}[h!]
\begin{center}
\begin{tabular}{| c| c |c|}
\hline
\multirow{3}{2.5cm}{{\centering\bf Noisy channel}}&\multirow{3}{2cm}{\centering\bf First family of invariants} &\multirow{3}{2.5cm}{{\bf Second and third families of invariants}}\\
 & & \\
 &  &\\
\hline\hline
\multirow{3}{2.5cm}{ Bit-flip channel }
&\multirow{3}{*} {$\langle\sigma_x\rangle$}& \multirow{3}{*}{$\dfrac{\langle \sigma_y\rangle}{\langle\sigma_z\rangle}$} \\
&&\\
& &  \\ 
\hline
  \multirow{3}{2.5cm}{Phase-flip channel} & \multirow{3}{*}{$\langle\sigma_z\rangle$} & \multirow{3}{*}{$\dfrac{\langle \sigma_x\rangle}{\langle\sigma_y\rangle}$}\\ 
  &&\\
  & &\\\hline
\multirow{3}{2.5cm} {Combination of bit and phase-flip channel}  & \multirow{2}{*}{$\langle\sigma_y\rangle$} & \multirow{3}{*}{$\dfrac{\langle \sigma_x\rangle}{\langle\sigma_z\rangle}$}\\
&&\\
 &&\\\hline 
 \multirow{3}{2.5cm}{Dephasing channel}&\multirow{3}{*}{$\langle \sigma_z\rangle$}  & \multirow{3}{*}{$\dfrac{\langle \sigma_x\rangle}{\langle \sigma_y\rangle}$}\\
&&\\
&&\\\hline
 \multirow{3}{2.5cm}{ Depolarizing channel}
& \multirow{3}{*}{--} &  \multirow{3}{2cm}{~~~~$\dfrac{\langle \sigma_x\rangle}{\langle\sigma_z\rangle}, \dfrac{\langle \sigma_y\rangle}{\langle\sigma_z\rangle}$}\\
&&\\
&&\\\hline
\multirow{3}{2.5cm} {Equiprobable bit and phase-flip channel}  & \multirow{2}{*}{--} & \multirow{3}{*}{$\dfrac{\langle \sigma_x\rangle}{\langle\sigma_z\rangle}$}\\
&&\\
 &&\\\hline 
 \multirow{3}{2.5cm}{ ADC}
& \multirow{3}{*}{--} & \multirow{3}{*}{$\dfrac{\langle \sigma_x\rangle}{\langle\sigma_y \rangle}$,~{\bf \boldmath$\dfrac{\langle\sigma_x\rangle\langle \sigma_y\rangle}{\langle\pi_z^-\rangle}$}}\\
&&\\
&&\\\hline
\multirow{3}{2.5cm}{ GADC}
& \multirow{3}{*}{--} & \multirow{3}{*}{$\dfrac{\langle \sigma_x\rangle}{\langle\sigma_z\rangle}$}\\
&&\\
&&\\\hline
 \end{tabular}
\caption{ Sets of invariants for various noisy channels of a qubit. The invariant quantity written in the boldface belongs to the third family of invariants.}
    \label{tab:qubit}
    \end{center}
\end{table}

The complete summary of the invariants of a qubit and a quNit is given in tables (\ref{tab:qubit}) and (\ref{tab:quNit}) respectively for various noisy channels. It is clear from table (\ref{tab:qubit}) that there are many channels for which no invariant is obtained from the first family. However, the second family provides invariants for all the channels. Only amplitude--damping channel has invariant from the third family.

Table (\ref{tab:quNit}) shows that the invariants belonging to the first family are rather limited in number in comparison to the second family. In fact, for some noisy channels, while no invariant belongs to the first family, the second family contains a large number of independent invariants. The third family provides an invariant for an amplitude--damping channel. The larger number of invariants that we have found for a quNit further strengthens the fact that larger information can be transferred with higher--dimensional states.
\begin{table}[h!]
\begin{center}
    \begin{tabular}{|c|c|c|c|c||c|}
    \hline
    \multirow{3}{2cm}{{\bf Noisy channel}}&\multirow{2}{1.5cm}{{\bf First family of invariants }}&\multirow{2}{1.5cm}{{\bf Number of Invariants}} & \multirow{3}{2cm}{{\bf Second and third families of invariants}}&\multirow{3}{1.5cm}{{\bf Number of invariants}}&\multirow{3}{1.5cm}{{\bf Total number of invariants}} \\
 & &&&&\\
 &  && &&
 \\
 &&&&&\\
 \hline\hline
        \multirow{3}{2cm}{Generalised flip error}& \multirow{3}{*}{$\langle X^m\rangle$} &\multirow{3}{*}{$N-1$}& \multirow{3}{*}{$\dfrac{\langle Z^m\rangle}{\langle(XZ)^m\rangle},\dfrac{\langle Z^m\rangle}{\langle Z^m X^l\rangle}$}& \multirow{3}{*}{$(N-1)^2$}&\multirow{3}{*}{$N(N-1)$}\\
        &&&&&\\
        &&&&&\\\hline
         \multirow{3}{2cm}{Generalised phase-error }& \multirow{2}{*}{$\langle Z^m\rangle$}&\multirow{3}{*}{$N-1$} & \multirow{3}{*}{$\dfrac{\langle X^m\rangle}{\langle(XZ)^m\rangle},\dfrac{\langle X^m\rangle}{\langle X^mZ^l\rangle}$}&\multirow{3}{*}{$(N-1)^2$}&\multirow{3}{*}{$N(N-1)$}\\
         &&&&&\\
         &&&&&\\\hline
          \multirow{6}{2cm}{Generalised combined flip and phase errors}& \multirow{6}{*}{$\langle (XZ)^m\rangle$}&\multirow{6}{*}{$N-1$} &\multirow{3}{*}{$\dfrac{\langle X^m\rangle}{\langle Z^{N-m}\rangle},~\dfrac{\langle X^m\rangle}{\langle X^lZ^{l-m}\rangle}$}&\multirow{6}{*}{$(N-1)^2$}&\multirow{6}{*}{$N(N-1)$}\\
          &&&&&\\
           &&&&&\\\cline{4-4}
           &&&\multirow{3}{*}{$\dfrac{\langle Z^m\rangle}{\langle Z^l X^{l-m}\rangle}$}&&\\
          &&&&&\\
         
        &&&&&\\\hline
                \multirow{3}{2cm}{Dephasing channel}& \multirow{3}{*}{$\langle D^{(kk+1)}\rangle$}&\multirow{3}{*}{$N-1$}&\multirow{3}{*}{$\dfrac{\langle S^{(kl)}\rangle}{\langle A^{(kl)}\rangle}$}&\multirow{3}{*}{$\dfrac{N(N-1)}{2}$}&\multirow{3}{*}{$\frac{(N-1)(N+2)}{2}$}\\
        &&&&&\\
        &&&&&\\\hline
         \multirow{6}{2cm}{Depolarizing channel}& \multirow{6}{*}{--}& \multirow{6}{*}{--}& \multirow{3}{*}{$\dfrac{\langle S^{(kl)}\rangle}{\langle D^{(kl)}\rangle},~\dfrac{\langle A^{(kl)}\rangle}{\langle D^{(kl)}\rangle}$,}&\multirow{6}{*}{$N^2-2$}&\multirow{6}{*}{$N^2-2$}\\
                 &&&&&\\
        &&&&&\\\cline{4-4}
        &&&\multirow{3}{*}{$\dfrac{\langle D^{(mn)}\rangle}{\langle D^{(ml)}\rangle}$}&&\\
        &&&&&\\
        &&&&&\\\hline
         \multirow{6}{2cm}{Flip error due to all transpositions}&\multirow{6}{*}{$\langle O\rangle$ }&\multirow{6}{*}{$1$}&\multirow{3}{*}{$\dfrac{\langle D^{(kl)}\rangle}{\langle D^{(k~N-1)}\rangle}$, $\dfrac{\langle O_2^{(k)}\rangle}{\langle D^{(k~N-1)}\rangle}$,}&\multirow{6}{*}{$\Big[\dfrac{5N}{2}\Big]-3$}&\multirow{6}{*}{$\Big[\dfrac{5N}{2}\Big]-2$}\\&&&&&\\
        &&&&&\\
        &&&\multirow{3}{*}{$\dfrac{\langle O_3^{(k)}\rangle}{\langle D^{(k~N-1)}\rangle}$}&&\\
        &&&&&\\
        &&&&&\\\hline
           \multirow{6}{2cm}{ADC }& \multirow{6}{*}{--}&\multirow{6}{*}{--}&\multirow{3}{*}{$\dfrac{\langle S^{(kl)}\rangle}{\langle A^{(kl)}\rangle}$,}&\multirow{3}{*}{ $\dfrac{N(N-1)}{2}$}&\multirow{6}{*}{ $\dfrac{N(N-1)}{2}+1$}\\
        &&&&&\\
        &&&&&\\\cline{4-5}
        &&&\multirow{3}{*}{\boldmath $\dfrac{\langle S^{(N-1~0)}\rangle\langle A^{(N-1~0)}\rangle}{\langle \pi_{N-1} \rangle}$}&\multirow{3}{*}{1}&\\
        &&&&&\\
        &&&&&\\\hline
        \multirow{3}{2cm}{GADC } & \multirow{3}{*}{--}&\multirow{3}{*}{--}& \multirow{3}{*}{$\dfrac{\langle S^{(kl)}\rangle}{\langle A^{(kl)}\rangle}$}&\multirow{3}{*}{$\dfrac{N(N-1)}{2}$}&\multirow{3}{*}{$\dfrac{N(N-1)}{2}$}\\
        &&&&&\\
        &&&&&\\\hline
         \end{tabular}
\end{center}
\caption{ The sets of invariants for various noisy channels of a quNit. The invariant written in the boldface belongs to the third family of invariants.}
    \label{tab:quNit}
\end{table}
Table (\ref{tab:quNit}) also gives insight into the relative impact of noisy channels. For example, in the channels corresponding to the generalised flip error, phase error, and combination of flip and phase error, the loss of information is $O(N)$. Whereas, in the channels such as dephasing, ADC, and GADC, the loss of information is $O\big(\frac{N^2}{2}\big)$. The maximum loss $O(N^2)$ of information occurs in the noisy channel in which error can be any of the transpositions.  

In table (\ref{tab:quNit}), except for the first three channels, the invariants are written directly in the basis operators defined in equations (\ref{eq:off-diagonal}) and (\ref{eq:sum_diagonal}). Hence, prescription for determining these via number count is straightforward. On the other hand, for the channels in the first three rows, the invariants are expectation values of unitary operators. This is not a hindrance as they may also be expressed in same basis operators. Explicitly, 
\begin{align}
   &X^m=\sum_{l=0}^{N-1} \frac{S^{(l+m~l)}+i A^{(l+m~l)}}{2},\quad ~~~ Z^m=\sum_{l=0}^{N-1} \omega^{(lm)}~d^{(l)}, \nonumber\\
   &(XZ)^m=\sum_{l=0}^{N-1}\omega^{ml+m} \frac{S^{(l+m~l)}+i A^{(l+m~l)}}{2},~~X^mZ^n=\sum_{l=0}^{N-1} \omega^{nl}\frac{S^{(l+m~l)}+i A^{(l+m~l)}}{2}.
\end{align}
 Thus, the task of retrieval of encoded information reduces to the task of finding expectation values of basis operators.

The expectation value of the diagonal basis operators is simply the population in that level. However, the expectation values of symmetric and anti-symmetric basis operators are the relative population in their eigenstates. That is,
\begin{align}
\langle d^{(k)}\rangle=\langle \pi_{k}\rangle,~~\langle S^{(kl)}\rangle= \langle\pi_{kl}^+\rangle-\langle\pi_{kl}^-\rangle,~~\langle A^{(kl)}\rangle= \langle\pi_{kl}^{+i}\rangle-\langle\pi_{kl}^{-i}\rangle,
\end{align}
where $\pi_{k}$ represents the projection operator in the states $\ket{k}$. Similarly, projection operators $\pi_{kl}^{\pm}$ and $\pi_{kl}^{\pm i}$ correspond to projection operators for the states $|kl^{\pm}\rangle$ and $|kl^{\pm i}\rangle$ respectively. These states are defined as:
\begin{equation}
   |kl^{\pm}\rangle=\dfrac{|k\rangle\pm |l\rangle}{\sqrt{2}},~~~~~~ |kl^{\pm i}\rangle=\dfrac{|k\rangle\pm i|l\rangle}{\sqrt{2}}.
\end{equation}
In this manner, by performing measurements in the appropriate basis states, expectation values of operators and hence information encoded in invariants can be retrieved.
If photonic sources of light are used, the expectation value of the projection operator is equal to the count rate in the state whereas, for a classical source of light, it is the intensity in that mode.


\section{Conclusion}
\label{conclusion}
In summary, we have laid down a formalism to extract invariants for a number of noisy channels. The information encoded in these quantities gets transferred without any error. This scheme of transferring error-free information alleviates the need for an entangled state, whose preparation is a difficult task. It also works for mixed states, which further reduces the burden of preparation of pure states.

This work opens up a number of possibilities. For example, the transfer of information using this formalism for bi-partite and multi-party systems constitutes an interesting study.  Multi-party states may also be employed to transfer error-free information to multiple users under noisy communication channels. Additionally, the formalism proposed in this work can also be used to transfer error-free information in some restricted scenarios. For example, the information may be encoded in the invariants in a way that it can be retrieved only if all the parties collaborate. 

 \acknowledgement
 It is a pleasure to thank 
 Sooryansh Asthana for several discussions, various insights, and help in improving the quality of the presentation. Rajni thanks UGC for funding her research.

\section*{Author Contribution Statement}
Both the authors have contributed equally in all respects.


\begin{appendix}
\section{Invariant quantities for a bit-flip channel}
\label{app:bit-flip}
A state $\rho$ after passing through a bit-flip channel changes to:
\begin{equation}
\rho\rightarrow\rho'=(1-p)\rho+p~\sigma_x\rho~  \sigma_x.
\end{equation}
The two quantities which can be used for encoding information for error-free transmissions are:
\begin{equation}
    \mathcal{I}_1=\langle \sigma_x\rangle,~~~~\mathcal{I}_2=\frac{\langle \sigma_y\rangle}{\langle \sigma_z\rangle}.
\end{equation}
The invariance of these two quantities can be seen as follows:
\begin{align}
 &\mathcal{I}_1=\langle \sigma_x\rangle_{\rho'}=\rho'_{01}+\rho'_{10}~=~\rho_{01}+\rho_{10}~=~ \langle \sigma_x\rangle_{\rho}\\
 &\mathcal{I}_2=\frac{\langle \sigma_y\rangle_{\rho'}}{\langle \sigma_z\rangle_{\rho'}}=\dfrac{i(\rho'_{01}-\rho'_{10})}{\rho'_{00}-\rho'_{11}}=\frac{(1-2p)i(\rho_{01}-\rho_{10})}{(1-2p)(\rho_{00}-\rho_{11})}\nonumber\\
 &~~~~~~~~~~~~~~~=~\frac{i(\rho_{01}-\rho_{10})}{\rho_{00}-\rho_{11}}~=~\frac{\langle \sigma_y\rangle_{\rho}}{\langle \sigma_z\rangle_{\rho}}.
\end{align}
\section{Invariant quantities for a generalised flip channel}\label{app:quNit_flip}
An $N-$ dimensional state $\rho$ after passing through generalised flip channel changes to:
 \begin{equation}\label{eq:quditflip}
     \rho\rightarrow\rho'= \sum_{r=0}^{N-1} p_r X^r\rho~ (X^r)^\dagger ,~~~~~~~~~~, 0\leq p_r\leq 1,~\sum_{r=0}^{N-1}p_r=1,
 \end{equation}
 where $X=\sum_{k=0}^{N-1} \ket{k+1}\bra{k}$ . The three sets of invariant quantities, for this noisy channels are listed as:
 \begin{equation}
     \mathcal{I}^{(m)}_{1}=\langle X^m\rangle ,~~~\mathcal{I}_{2}^{(m)}=\frac{\langle Z^m\rangle}{\langle (XZ)^m\rangle},~~~\mathcal{I}_{2}^{(mnl)}=\frac{\langle Z^mX^n\rangle}{\langle Z^mX^l\rangle},~~~~m\neq n\neq l,
 \end{equation}
 where $ Z= \sum_{k=0}^{N-1} \omega^k \ket{k}\bra{k}$ and $\omega= e^{\frac{2\pi i}{N}}$.
 To show invariance of these quantities,  the three properties namely, (i) invariance of trace under unitary operations, (ii) Commutation relation of an operator $A$ with some function of itself, i.e.,$[A,f(A)]=0$, and (iii) $Z^{n_1}X^{n_2}=\omega^{n_1n_2} X^{n_2}Z^{n_1}$ are used.\\
 \noindent{\bf First set of quantities:} Let's first consider operator $X^m$,
 \begin{equation}
     \langle X^m\rangle_{\rho'={\rm Tr}(X^m\rho')}={\rm Tr} \Big( X^m\sum_{r=0}^{N-1} p_r X^r\rho (X^r)^\dagger \Big)=\sum_{r=0}^{N-1} p_r {\rm Tr}\Big( X^m X^r\rho~ (X^r)^\dagger\Big).
 \end{equation}
 
 Since the operator $X^m$ commutes with all the powers of $X$, the above equation changes to:
 \begin{eqnarray}
    && \langle X^m\rangle_{\rho'}= \sum_{r=0}^{N-1} p_r {\rm Tr}\Big( X^r X^m\rho (X^r)^\dagger\Big)= \sum_{r=0}^{N-1} p_r {\rm Tr}  \Big( X^m\rho\Big)={\rm Tr}  \big( X^m\rho\big),\nonumber\\
    && \langle X^m\rangle_{\rho'}=\langle X^m\rangle_{\rho}
 \end{eqnarray}
 \noindent{\bf Second set of quantities:} Let's first calculate expectation value of $Z^m$,
 \begin{equation}
     \langle Z^m\rangle_{\rho'}={\rm Tr}\big(Z^m\rho'\big)={\rm Tr} \Big( Z^m\sum_{r=0}^{N-1} p_r X^r\rho~ (X^r)^\dagger \Big)=\sum_{r=0}^{N-1} p_r {\rm Tr}\Big( Z^m X^r\rho~ (X^r)^\dagger\Big).
 \end{equation}
 Using the relation, $Z^{n_1}X^{n_2}=\omega^{n_1n_2} X^{n_2}Z^{n_1}$,
  \begin{eqnarray}\label{eq:Z}
    && \langle Z^m\rangle_{\rho'}= \sum_{r=0}^{N-1} p_r {\rm Tr}\Big( \omega^{rm} X^r Z^m\rho~ (X^r)^\dagger\Big)= \sum_{r=0}^{N-1} p_r \omega^{rm}{\rm Tr}  \Big( Z^m\rho\Big),\nonumber\\
    && \langle Z^m\rangle_{\rho'}=\sum_{r=0}^{N-1} p_r ~\omega^{rm}\langle Z^m\rangle_{\rho}.
 \end{eqnarray}
 Now, calculating expectation value of $(XZ)^m$,
 \begin{equation}
     \langle (XZ)^m\rangle_{\rho'}={\rm Tr}((XZ)^m\rho')={\rm Tr} \Big( (XZ)^m\sum_{r=0}^{N-1} p_r X^r\rho~ (X^r)^\dagger \Big)=\sum_{r=0}^{N-1} p_r {\rm Tr}\Big( (XZ)^m X^r\rho (X^r)^\dagger\Big).
 \end{equation}
 Using the relation, $Z^{n_1}X^{n_2}=\omega^{n_1n_2} X^{n_2}Z^{n_1}$,
  \begin{eqnarray}\label{eq:XZ}
    && \langle (XZ)^m\rangle_{\rho'}= \sum_{r=0}^{N-1} p_r {\rm Tr}\Big( \omega^{rm} X^r (XZ)^m\rho (X^r)^\dagger\Big)= \sum_{r=0}^{N-1} p_r ~\omega^{rm}{\rm Tr}  \Big( (XZ)^m\rho\Big),\nonumber\\
    && \langle (XZ)^m\rangle_{\rho'}=\sum_{r=0}^{N-1} p_r ~\omega^{rm}\langle (XZ)^m\rangle_{\rho}
 \end{eqnarray}
  Equations (\ref{eq:Z}) and (\ref{eq:XZ}) clearly indicate that the ratio of the two remain invariant under the noisy evolution of a state, i.e.,
  \begin{equation}
      \frac{\langle Z^m\rangle_{\rho'}}{\langle (XZ)^m\rangle_{\rho'}}=\frac{\langle Z^m\rangle_{\rho}}{\langle (XZ)^m\rangle_{\rho}}.
  \end{equation}
  \noindent{\bf Third set of quantities:}
For this set of invariants, one need to calculate expectation value of operator $Z^mX^n $,
\begin{eqnarray}\label{eq:mn}
     &&\langle Z^mX^n\rangle_{\rho'}={\rm Tr}\big(Z^mX^n\rho'\big)={\rm Tr} \Big( Z^mX^n\sum_{r=0}^{N-1} p_r X^r\rho (X^r)^\dagger \Big)=\sum_{r=0}^{N-1} p_r {\rm Tr}\Big( Z^mX^n X^r\rho (X^r)^\dagger\Big),\nonumber\\
     &&=\sum_{r=0}^{N-1} p_r {\rm Tr}\Big( Z^mX^r X^n\rho (X^r)^\dagger\Big)=\sum_{r=0}^{N-1} p_r {\rm Tr}\Big( \omega^{rm}X^rZ^m X^n\rho (X^r)^\dagger\Big)\nonumber\\
     &&=\sum_{r=0}^{N-1} p_r~ \omega^{rm} ~{\rm Tr}\Big( X^rZ^m X^n\rho (X^r)^\dagger\Big)=\sum_{r=0}^{N-1} p_r ~\omega^{rm} {\rm Tr}\Big( Z^m X^n\rho\Big)=\sum_{r=0}^{N-1} p_r~ \omega^{rm}\langle Z^mX^n\rangle_{\rho}.
\end{eqnarray}
Similarly, the expectation value of $Z^mX^l$ can be written as:
\begin{equation}\label{eq:ml}
    \langle Z^mX^l\rangle_{\rho'}=\sum_{r=0}^{N-1} p_r~ \omega^{rm}\langle Z^mX^l\rangle_{\rho}.
\end{equation}
Taking the ratio of the equations (\ref{eq:mn}) and (\ref{eq:ml}), one get:
\begin{equation}
    \frac{\langle Z^mX^n\rangle_{\rho'}}{\langle Z^mX^l\rangle_{\rho'}}=\frac{\langle Z^mX^n\rangle_{\rho}}{\langle Z^mX^l\rangle_{\rho} },
\end{equation}
which is an invariant quantity. Thus, we have shown that all three sets of quantities remain invariant under the noisy evolution of a state.
\end{appendix}

\end{document}